\documentclass[manuscript,screen,]{acmart}
 \usepackage{graphicx}
\usepackage[english]{babel}
\usepackage[utf8]{inputenc}
\usepackage{natbib}
\usepackage{lineno}

\title[Submitted for Publication in ACM Special Issue on Distributed Intelligence on the Internet]{Building Blocks to Empower Cognitive Internet with Hybrid Edge Cloud}

\author{Fay Arjomandi}
\email{Fay.Arjomandi@mimik.com}
\affiliation{%
    \institution{mimik}
        \country {USA}}
\author{Siavash Alamouti}
\email{Siavash.Alamouti@mimik.com}
\affiliation{%
  \institution{mimik}
  \country{USA}}
\author{Michel Burger}
\email{Michel.Burger@mimik.com}
\affiliation{%
  \institution{mimik}
    \country{USA}}
\author{Dr. Bashar Altakrouri}
\email{Bashar.Altakrouri@aramco.com}
\affiliation{%
    \institution{Aramco}
     \country{Saudi Arabia}}

\copyrightyear{2024}
\acmYear{2024}
\acmDOI{XXXXXXX.XXXXXXX}

\begin{document}
\nolinenumbers



\begin{abstract}
As we transition from the mobile internet era into the epoch of the 'Cognitive Internet,' a profound transformation is underway in the way we interact with technology, data, and intelligence. This article contends that the Cognitive Internet transcends the Cognitive Internet of Things (Cognitive IoT) paradigm, wherein connected objects independently acquire knowledge, cognitive abilities, and understanding of both physical and social realms. The Cognitive Internet emphasizes a shift towards an integrated intelligence and cognition of objects, things and systems that operate dynamically and within and across heterogeneous ecosystems and domains. 

To achieve its full potential and to distinguish itself from the Mobile Internet and Cognitive IoT, the Cognitive Internet inherently infuses intelligence both horizontally and vertically throughout the entire network allowing to fuse organically and fully the realm of cognitive IoT and human intelligence. This inherent intelligence empowers cognitive choreography, facilitating interactions between connected devices, services, entities, and individuals across diverse domains and industries. Importantly, it upholds the autonomy of decision-making and accommodates heterogeneous identities.

This paper explores the foundational elements, distinguishing characteristics, benefits, and the transformative industrial impact of the 'Cognitive Internet' paradigm. It emphasizes the pivotal role of resilient and adaptable AI infrastructures, underpinned by the innovation of hybrid edge cloud (HEC) platforms. This change in thinking heralds a new era, marked by the proliferation of cognitive services, the emergence of a Knowledge as a Service (KaaS) economy, heightened decision-making autonomy, sustainable digital progress, the evolution of data management and processing techniques, and an increased emphasis on privacy.

In its essence, this paper serves as an indispensable guide for comprehending and harnessing the transformative potential inherent in the Cognitive Internet. Elaborate case studies, forward-looking perspectives, and real-world applications further bolster the discourse, offering comprehensive insights into this emergent paradigm. 

\end{abstract}

\ccsdesc[500]{Computer systems organization}
\ccsdesc[500]{Software and its engineering}
\ccsdesc[500]{Security and privacy}
\ccsdesc[500]{Human-centered computing}
\ccsdesc[500]{Computing methodologies}

\keywords{AI, Automation, Autonomous, Decision-making, edge AI, Hybrid edge cloud, edge-engine, cognitive internet,   cognitive IoT,   Industrial automation,   offshore Autonomous operations,   zero trust security,   API gateway,   Context,   microservices,   offline-first,   context-aware systems,   ad-hoc elastic systems,   energy efficiency,   privacy-first,   zero-touch configuration }


\maketitle

\section{Introduction}
Can you imagine a world where our digital systems not only process information but learn from it, adapt to our needs, and autonomously drive actions? This is not a mere fantasy or a scene from a sci-fi movie; it is the reality of the rapidly emerging 'Cognitive Internet' era where all devices from consumer gadgets, industrial and medical devices to cars and trucks will have compute resources with local intelligence. Driven by AI, automation, and autonomous decision-making, we are experiencing a technological revolution that will dismantle application silos, paving the way for seamless, dynamic communication between near-autonomous devices and systems.

The Cognitive Internet extends beyond isolated apps and traditional SaaS models. The burgeoning push for automation lays the groundwork for an intelligent service ecosystem, where services running on smart devices such as robots, drones, smart appliances, and things are not just connected—they are 'cognitive.' They can discover abilities and exchange knowledge to collaborate as needed based on the context. They can discover and adapt to tasks much like the human brain, marking the rise of a new era of interconnectedness.

The Cognitive Internet is a core enabler for a Knowledge as a Service (KaaS) economy \cite{Abdullah}. It will allow industry to go beyond traditional and cognitive IoT \cite{Wu} where sensors and machines are single-function and static dumb devices with all the smarts in the cloud that cannot directly communicate or collaborate with other devices. Autonomy requires these devices to have local intelligence and be able to interact with other devices and systems. These devices leverage AI for knowledge and insights. In the Cognitive Internet era, intelligent devices and systems can understand, learn from, and adapt to evolving needs, thereby revolutionizing our digital experiences not just to mimic single-function physical interactions but to evolve and adapt based on context and knowledge.

The proliferation of smart devices—with enhanced computing capabilities, connectivity, and sensors like cameras, lidar, and radars—has led to an unprecedented volume of real-time data. Traditionally, we have utilized cloud data analytics to process this information, extract insights, and guide decision-making. However, the pervasive integration of AI is revolutionizing this approach. AI does not merely process data; it learns from it, improves algorithms, and autonomously drives actions with unprecedented precision and efficiency. This shift propels us into a future where data-driven insights are the starting point of an iterative, intelligent process that constantly evolves and adapts.

In response to this significant shift towards the Cognitive Internet, there is a growing necessity for a robust, flexible AI infrastructure—an infrastructure capable of harnessing the exponentially increasing real-time data volume and swiftly translating it into actionable insights. Given the distributed and heterogeneous nature of these devices and systems, we need distributed runtime environments that can host workflows directly with minimal reliance on networks and central cloud. These devices must have significant offline capabilities, adapt their workflows to the task at hand, and seamlessly update their AI models and infer to evolving needs. They must be able to expose their capabilities to other devices for system-level operations and collaborate with others. For instance, clusters of collaborating AI-enabled cameras, or a swarm of drones with different capabilities: custom embedded AI chips, GPUs, satellite connectivity, shared storage, etc.

In other words, smart AI-enabled objects should always offer some basic local cloud-server capabilities, including an API gateway, a runtime environment for microservices, local and global discovery, and microservice-level communications with other objects and systems and central cloud. We claim that these capabilities can be best provided by Hybrid Edge Cloud (HEC) \cite{Siavash}  as illustrated in Figure 1.
HEC provides the essential scalability, flexibility, and reduced latency achieved by processing data at the source, increased bandwidth efficiency by minimizing data transfers to the cloud, improved data privacy and security from retaining sensitive information at the source, and local decision-making autonomy. Moreover, it minimizes energy consumption of the data path. All these factors underscore the essential role of HEC in the Cognitive Internet era.

For instance, consider an auto manufacturer planning to launch a fleet of autonomous vehicles. Each vehicle generates massive amounts of data from onboard sensors, cameras, and Lidar. By processing this data locally and in real-time using HEC, the vehicles can make immediate, life-critical decisions, while still sending selective data back to the central cloud for long-term learning and refinements and updates from global training and learning. HEC will allow the car to start logging events, understand the context and act immediately from its inception (even when partially assembled) throughout its full life cycle due to the rich surrounding cognitive ecosystem.

Adopting HEC as a distributed cloud platform for AI-powered (i.e., cognitive) solutions is a transformative step toward realizing the full potential of the Cognitive Internet. This shift strengthens our capacity to efficiently process and utilize data, powers cognitive services, and lays the foundation for a more responsive, secure, and intelligent digital world.

In this paper, we will delve deeper into the building blocks and concepts of Cognitive Internet, explore the benefits and challenges of this transformation, and provide guidance on navigating this new landscape with the focus on HEC. Our goal is to equip decision-makers with the understanding and tools necessary to harness the potential of the Cognitive Internet and the accompanying infrastructure advancements.  

\begin{figure}[ht]
    \centering
    \includegraphics[width=.65\linewidth]{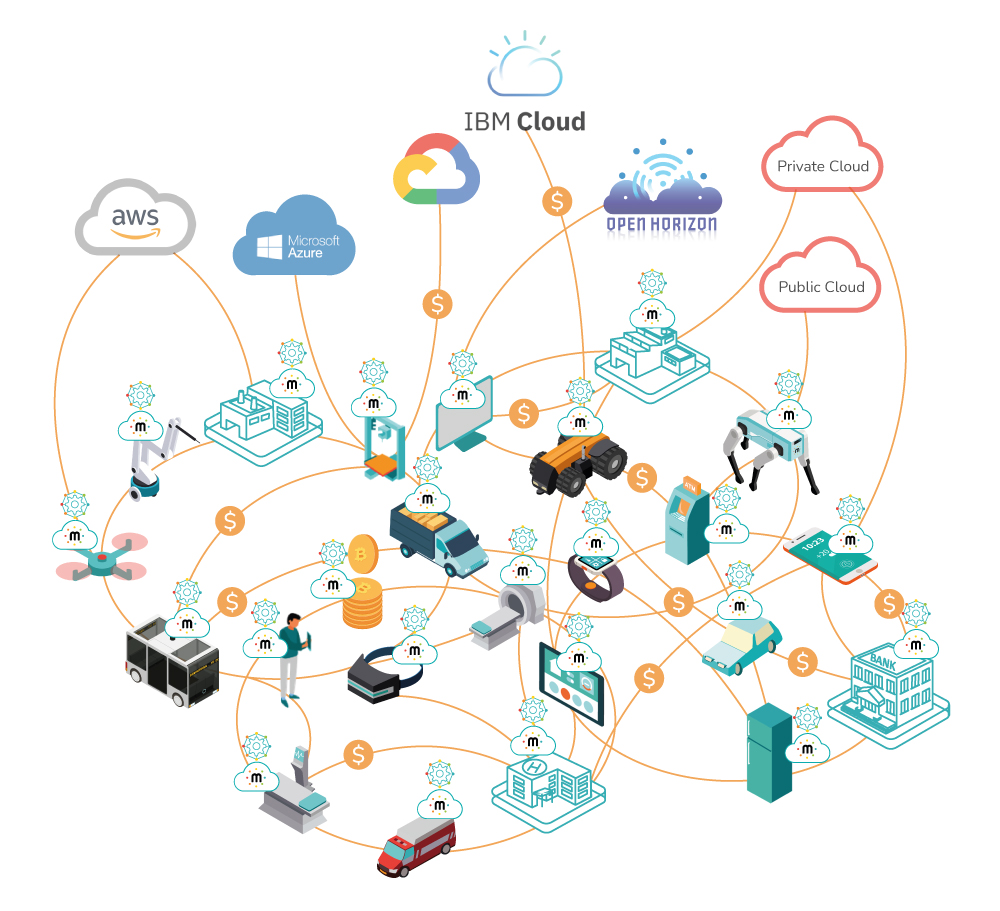}
    \caption{High Level Illustration of Hybrid Edge Cloud (HEC)}
    \label{fig:enter-label}
\end{figure}

\section{EVOLUTION TO COGNITIVE INTERNET}
In the Cognitive Internet era, where hundreds of billions of smart devices, apps, and processes stand poised to transform virtually every industry through automation, it’s essential that digital solutions embody the intricate interactions prevalent in the physical space. As the digital realm increasingly converges, the boundary between them grows ever thinner. Previously, digital models were constructed to mimic reality and were continuously fed by data to maintain their accuracy. However, a model, no matter how sophisticated and reality, has always been two distinct entities. The Cognitive Internet seeks to bridge this gap, ensuring that digital systems do not just emulate but genuinely mirror and seamlessly integrate with this complex, interconnected reality.

\begin{table}[ht]
    \centering
   \caption{Mobile Internet vs. Cognitive IoT and Cognitive Internet}
   \label{tab:freq}
    \begin{tabular}{|c|c|c|c|} \hline
    \toprule
    Attribute & Mobile Internet & Cognitive IoT & Cognitive Internet\\
    \midrule
    {\footnotesize   Key Characteristics} &	{\footnotesize   Siloed apps and}	& {\footnotesize   Autonomous connected} &	{\footnotesize   Integrated intelligence} \\
    & {\footnotesize   cloud-hosted data} & {\footnotesize   objects} & {\footnotesize   and cognition}\\ \hline
{\footnotesize   Interaction and }&	{\footnotesize   Limited device interaction, }&	{\footnotesize   Connected objects}	& {\footnotesize   Dynamic, interconnected objects}\\
   {\footnotesize    Intelligence} &	{\footnotesize   mainly data consumption} &	{\footnotesize   with independent knowledge} &	{\footnotesize   with integrated intelligence}\\ \hline
    {\footnotesize   Autonomy} &	{\footnotesize   Limited autonomy; } &	{\footnotesize   Increasing autonomy; } &	{\footnotesize   High autonomy; local and }\\
    &	{\footnotesize   reliance on the cloud}	& {\footnotesize   local intelligence} &	{\footnotesize   distributed intelligence}\\ \hline
   {\footnotesize    Automation and }&	{\footnotesize   Limited automation, }	&{\footnotesize   Automation and collaboration}	& {\footnotesize   Automation, knowledge exchange, }\\ 
    {\footnotesize   Collaboration}	& {\footnotesize   isolated apps}	& {\footnotesize   among connected objects} &	{\footnotesize   and collaboration}\\ \hline
   {\footnotesize    Data Handling	}&{\footnotesize   Traditional cloud}	&{\footnotesize   Data analytics, } &	{\footnotesize   Real-time data processing }\\
   & {\footnotesize   data analytics}	& {\footnotesize   cloud-based intelligence}	& {\footnotesize   and local intelligence}\\ \hline
    {\footnotesize   AI Integration}	&{\footnotesize   Limited AI integration	}&{\footnotesize   AI for knowledge}	&{\footnotesize   Extensive AI integration, }\\
    && {\footnotesize   and insights}	& {\footnotesize   local and distributed AI}\\ \hline
    {\footnotesize   Infrastructure Requirements}	&{\footnotesize   Cloud-centric infrastructure}	&{\footnotesize   Cloud and local device }&	{\footnotesize   Hybrid Edge Cloud (HEC) }\\
    & & {\footnotesize   intelligence}&	{\footnotesize   for local and distributed AI}\\ \hline
    {\footnotesize   Challenges	}& {\footnotesize   Data privacy, latency,	}& {\footnotesize   Interoperability, data privacy }& {\footnotesize   Scalability, flexibility, and}\\
    & {\footnotesize    and cloud dependence}&	&	{\footnotesize   reduced latency}\\ \hline
    {\footnotesize   Examples}&	{\footnotesize   Traditional mobile apps}	&{\footnotesize   Smart devices with}	&{\footnotesize   Smart devices with}\\
    &	& {\footnotesize   basic intelligence}	&{\footnotesize   integrated intelligence}\\ \hline
    \bottomrule
    \end{tabular}
    
\end{table}

    During the previous Mobile Internet era, enterprises primarily relied on siloed, vertical mobile applications that accessed and consumed largely cloud-hosted data. As we transition into the Cognitive Internet era, the need for horizontal systems becomes more pressing. Such systems must transcend the challenges presented by silos, efficiently connecting all solution silos that underpin enterprise systems. The interchange of information should be seamless, triggered by actions that could lead to transactions, initiate business processes, or guide decisions.

    In the era of the Cognitive Internet, solutions transcend the limitations of the traditional Internet of Things (IoT), ushering in an era of the Internet of Systems. This interconnected network comprises subsystems and services that operate autonomously yet collaboratively, mirroring human-like behaviors in their interactions. However, the journey to achieve this level of digital infrastructure integration enterprises must address a multitude of obstacles, including:
    \begin{itemize}
        \item Maximizing the extraction of actionable insights from data to enable intelligent, knowledge-centric automation.
        \item Maintain maximal independence from proprietary solution stacks, specific cloud providers, and device compatibility constraints, including various operating systems and network technologies.
        \item Safeguarding sensitive enterprise, customer, and employee data while staying compliant with evolving data regulations.
        \item Enabling real-time adaptability, informed by the knowledge gained, to swiftly respond to the ever-changing environment and adjust device behavior accordingly.
        \item Ensure long-term cost efficiency and sustainability for the entire solution.
    \end{itemize}

\subsection{Predictive Maintenance for Manufacturing Equipment}

In a manufacturing plant, various machines and equipment are interconnected to optimize production. HEC enables any smart device with CPU resources to host microservices and communicate at microservice level with all other devices, expose and make available its resources to others, and create ad-hoc service mesh to complete tasks. In other words, each machine or equipment can host its own microservices for data collection, processing, and analysis. These microservices can continuously monitor sensors, collect data, analyze them locally, and drive actions based on data-driven decisions by themselves or with the collaboration of other machine or equipment.

When anomalies or potential issues are detected, the machines not only communicate directly with each other, with fieldworker devices, and local gateways in a peer-to-peer manner, but also collectively aggregate knowledge and coordinate actions among the respective local devices and things  For instance, if one machine detects a problem, it can notify adjacent machines to adjust their operations taking into account both the individual machine's context and knowledge, as well as the local context and knowledge of surrounding machines. However, for more complex predictive maintenance algorithms that require extensive historical data and advanced analytics, the machines can collaborate with central cloud-based services.

In this case, HEC provides the ability to perform localized decision-making when needed and tap into centralized cloud resources for more computationally intensive tasks, ensuring timely maintenance actions and optimized production.

Figure 2 shows a predictive maintenance use case where a field technician is notified of an anomaly by running a “field assist app”. The field assist microservice on the app fetches all necessary information from the two existing microservices running on the on-board computers on the equipment and addresses the issue.

\begin{figure}[ht]
    \centering
    \includegraphics[width=.75\linewidth]{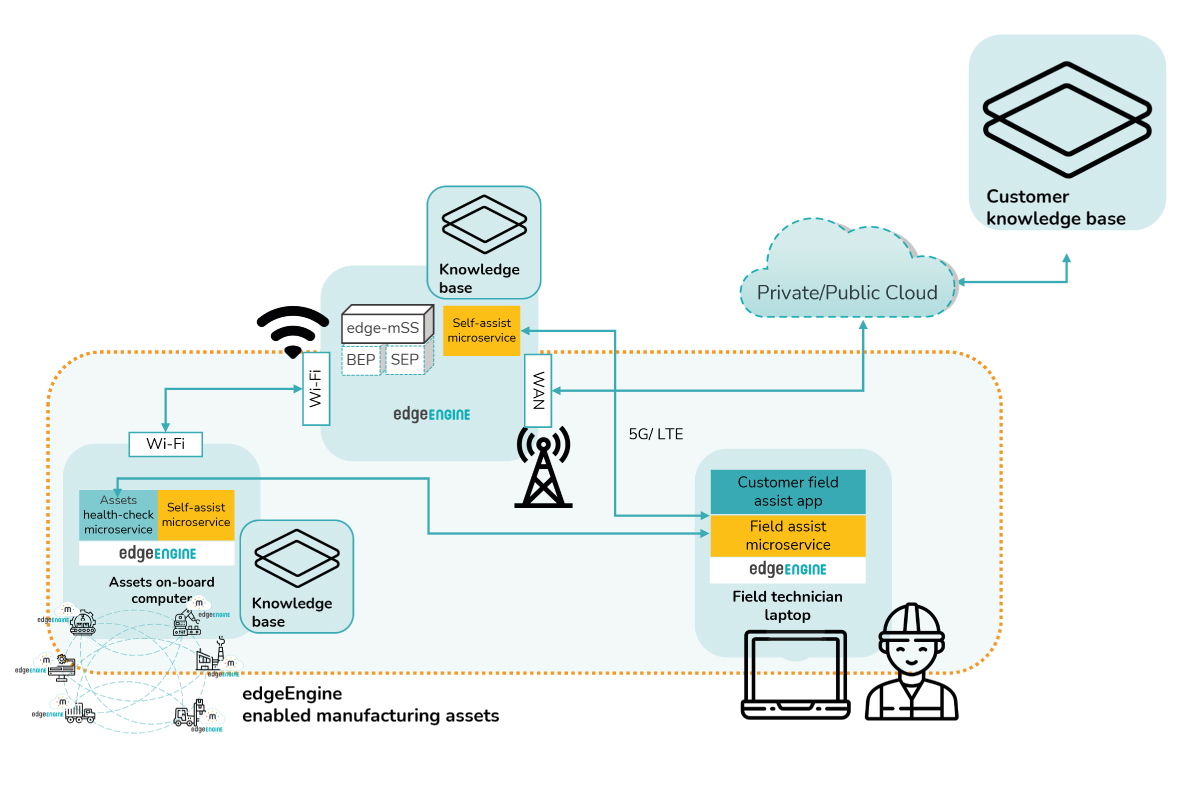}
    \caption{Industrial automation use case with HEC.}
\end{figure}

\subsection{Robotic Automation in Warehouse Logistics}
In a warehouse or logistics center, autonomous robots can host their own microservices for real-time data processing and decision making. As illustrated in Figure 3, these robots can communicate with each other directly to share information about their status and coordinate their movements, all without the need for central cloud resources.

For instance, a single robot can independently learn from its surroundings, adjust its behavior, and facilitate the same learning and adaptation process among neighboring robots, even if they come from different manufacturers. This peer-to-peer communication and collaborative adaptation not only improve navigation and safety but also exemplify the system's self-reliance, eliminating the need for central cloud dependency and highlighting the seamless integration of diverse robotic platforms.

Meanwhile, more complex tasks, such as optimizing the overall warehouse operations, can be coordinated through central cloud-based services. The robots can periodically upload data related to their activities, and the central cloud can analyze this data to make high-level decisions like optimizing robot paths, managing inventory levels, and adapting to changing warehouse conditions. This dual approach underscores the crucial balance that must exist between local decision-making—where quick, autonomous responses are vital—and more global, centralized decision-making that takes a broader view. Such systems, by dynamically adjusting to data over time, lean towards becoming 'eventually optimal,' ensuring consistent improvement and adaptability.

In both scenarios, the use of microservices hosted directly on machines and robots, combined with peer-to-peer communication via ad-hoc edge service mesh, enables real-time, context-driven decision making at the edge. The HEC architecture remains essential for uploading resource-intensive tasks to centralized cloud services, when necessary, thereby maintaining the flexibility and efficiency of industrial automation systems.

These examples highlight the myriad considerations that enterprises must address in the Cognitive Internet era, marking a pivotal point in the evolution of the Internet.
\begin{figure} [ht]
    \centering
    \includegraphics[width=.6\linewidth]{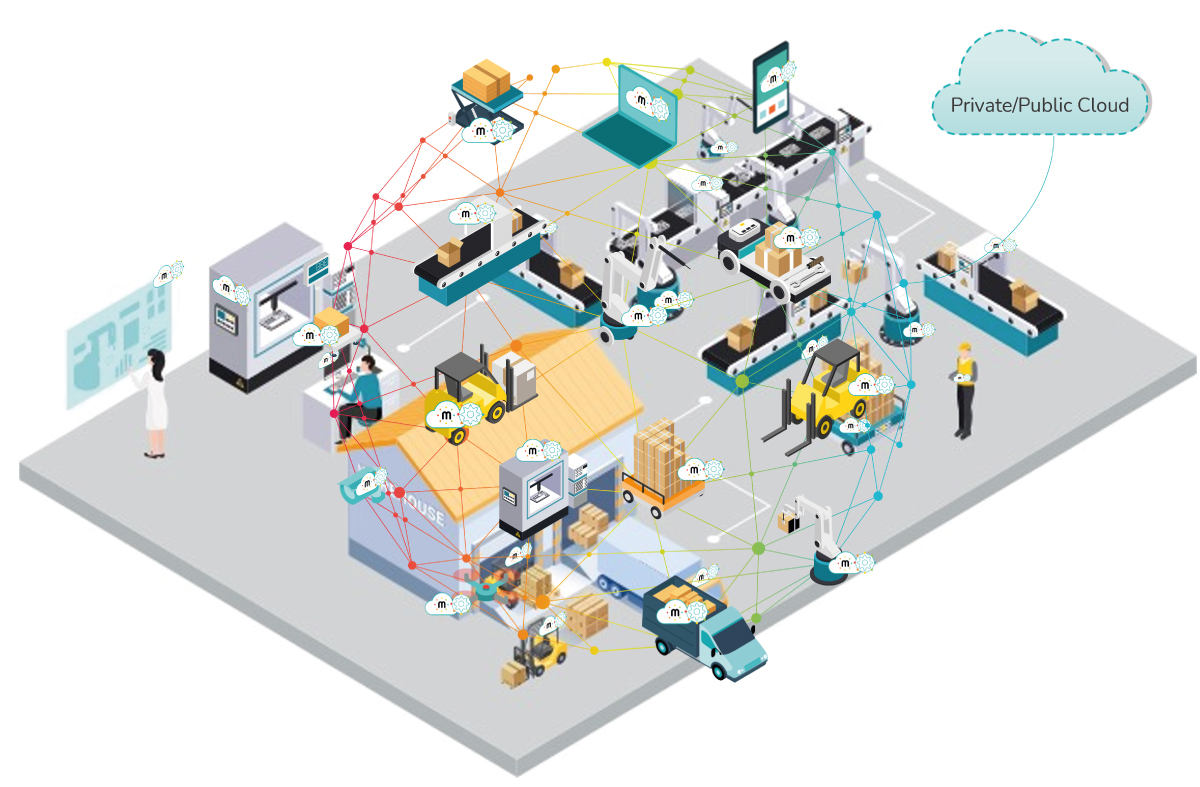}
    \caption{Automation use case example with HEC.}
\end{figure}

\section{DIGITAL TRANSFORMATION IN THE COGNITIVE INTERNET ERA}

In the era of the Cognitive Internet, digital transformation necessitates a holistic understanding of business processes that begins at end-point devices. End-point devices, in this context, encompass a rich diversity of hardware, ranging from smartphones to industrial machines, robots, drones, and various other intelligent devices. Each of these devices exhibits varying capabilities, connectivity options, power constraints, and geographical distribution.

This transformation journey must embrace and harness this diversity while redefining the dynamics between the end devices and the cloud. At one end of the spectrum, certain end devices and systems boast advanced data processing capabilities, enabling localized AI-driven decision-making. Conversely, the end-point ecosystem might also encompass simpler sensors and devices with limited processing prowess.

The essence of digital transformation lies in seamlessly integrating this spectrum of systems, crafting an Internet of Systems where processes operate autonomously and collaborate as required. Such a transformation necessitates a resilient, adaptable, scalable, secure, and privacy-conscious infrastructure that accommodates the inherent heterogeneity and diversity of end-point devices.

Conventional backend integration methods may prove impractical due to constraints tied to cost, latency, bandwidth, energy efficiency, or privacy concerns. In manufacturing, for instance, a production facility equipped with an array of automated machines and robots can benefit from this approach. Each machine and robot serve as an end device, equipped with local processing capabilities.

These end devices collaborate with each other autonomously to optimize production processes, monitor equipment health, and make real-time adjustments to ensure quality and efficiency. Simultaneously, they contribute data to a central cloud system for global learning and coordination. The central cloud can analyze historical data to predict maintenance needs, optimize production schedules, and enhance overall manufacturing operations.

This 'end-device-centric' approach to digital transformation, originating at the end devices and culminating in an Internet of Systems, represents a pivotal departure from traditional 'cloud-down' strategies. It underscores the importance of empowering autonomous devices such as machines, robots, and drones with local processing capabilities while simultaneously enabling global learning and coordination in all industries.
\section{UNDERSTANDING THE COGNITIVE INTERNET}
\subsection{Definition and Characteristics}
The Cognitive Internet, as the cornerstone of the new digital era, represents a transformative shift in how we perceive and interact with intelligent systems. It encompasses a realm where these systems not only collect and share data, as seen with the Internet of Things (IoT), but also possess the capability to comprehend, learn, decide, and take purposeful actions based on vast datasets.

To truly grasp the essence of the Cognitive Internet, let's draw a distinction between it and the IoT. In the IoT era, devices were elevated to 'smart' status by allowing them to gather and exchange data. However, the Cognitive Internet takes this concept a step further. Here, devices don't merely serve as data conduits; they become cognitive agents that process, interpret, and act upon the information they acquire.

\subsection{AI, Automation, and Autonomous Decision-Making}
At the heart of the Cognitive Internet lies the pivotal role of Artificial Intelligence (AI). Yet, it's essential to recognize that AI's role in this context differs significantly from its application in previous Internet iterations. In the IoT era, AI primarily found its place in predictive analytics. However, in the Cognitive Internet, AI assumes the role of the 'brain' that empowers systems to autonomously analyze data, make informed decisions, and dynamically adapt to evolving circumstances.

\subsection{Understanding Contextual Dimensions in the Cognitive Internet}
In the realm of the Cognitive Internet, understanding context is paramount. At any point in time, context is not just about the immediate surroundings; it delves deeper, answering questions like 'Who am I?', ‘What is my abilities’, 'Who do I serve?', 'What is my role?', 'Where am I?', and 'What's my relationship with others?'. This introspective understanding provides the situational backdrop against which systems operate, helping to tailor responses, optimize operations, and improve decision-making. To elucidate this, we identify several key contextual dimensions:
\textbf{} 
    \begin{itemize}
        \item \textbf{Temporal Context}: This pertains to the evolution of the system over time. It encompasses changes in the system's structure, software, or requirements. Such shifts might be driven by technological advancements, user demands, or external environment variations.
        \end{itemize}
    \begin{itemize}
        \item \textbf{Security Context}: 
        \begin{itemize}\item \textbf{Trust Levels}: Not all nodes within a system are equally trusted. This disparity becomes especially pronounced in decentralized settings like peer-to-peer systems or public blockchains where various actors have varying credibility.
        
            \item \textbf{Threat Model}: Recognizing potential threats is vital. The system's anticipated adversaries or types of attacks will significantly influence its design, calling for tailored defense mechanisms.
        \end{itemize}
        \end{itemize}
    \begin{itemize}
        \item \textbf{Hardware Context}: 
        \begin{itemize}
        \item\textbf{Device Types}: Distributed systems in the Cognitive Internet landscape may operate on a vast spectrum of devices, ranging from high-performance data center servers to more constrained IoT devices.
        \item\textbf{Network Conditions}: The nuances of the network play a role. Factors like reliability, speed, and the nature of the connection (be it wired or wireless) between nodes dictate the system's operational context.
        \end{itemize}
        \end{itemize}
    \begin{itemize}
        \item \textbf{Operational Context}: 
        \begin{itemize}
        \item\textbf{Load}: Systems must be designed to handle varying workloads. The quantity and type of tasks the system encounters could fluctuate extensively.
        \item\textbf{Resource Availability}: The availability of resources like memory, CPU, bandwidth, and storage can greatly influence system performance, resilience, and scalability.
       \item \textbf{Maintenance State}: At any given time, certain nodes might be offline due to updates, repairs, or other operational reasons. This can impact the system's overall availability and reliability.
          \end{itemize}
        \end{itemize}
    \begin{itemize}   
        \item \textbf{Data Context}: 
        \begin{itemize}
        \item\textbf{Data Cohesion} : Understanding the interrelation between data elements is crucial. Data frequently accessed in tandem might be stored jointly to expedite retrieval.
        \item\textbf{Data Locality} : Performance is optimized when data is stored proximate to its processing locus. This principle ensures reduced latency and more efficient data handling.
        \end{itemize}
        \end{itemize}
    \begin{itemize}      
        \item \textbf{Data Privacy Context}: 
        \begin{itemize}   
        \item\textbf{User Consent}: Understanding and managing permissions given by users for data collection, processing, and sharing.
       \item \textbf{Anonymization and Pseudonymization}: Techniques used to protect personal data, ensuring individuals remain unidentifiable.
        \item\textbf{Regulatory Compliance}: Adhering to data protection regulations such as GDPR, CCPA, etc., depending on the geographical location and nature of operations.
         \end{itemize}
        \end{itemize}
    \begin{itemize}       
        \item \textbf{Data Sovereignty Context}: 
        \begin{itemize}   
        \item\textbf{Data Residency}: The physical location where data is stored, often influenced by regulatory and compliance needs.
        \item\textbf{Cross-border Transfers}: Considerations and regulations around transferring data across country borders.
        \item\textbf{Local Regulations}: Recognizing and adhering to local laws governing data storage, processing, and transmission.
        \end{itemize}
        \end{itemize}
        \begin{itemize}   
        \item \textbf{Interactivity Context} : 
        \begin{itemize}   
        \item\textbf{Node Relations}: In a connected ecosystem like the Cognitive Internet, understanding how nodes communicate and collaborate is crucial. Whether they function in hierarchies, as peers, or in more complex relationships, this context shapes the flow of information and decision-making processes. 
        \item\textbf{External Interfaces}: Not only is the interaction between nodes within the system vital, but so is the system's interaction with external entities. This might involve interfacing with other networks, systems, or even with users, and understanding these interactions is key to optimizing system responsiveness and efficiency.
        \end{itemize}
        \end{itemize}
        \begin{itemize}   
        \item \textbf{Environmental Context}: 
        \begin{itemize}   
        \item\textbf{Physical Conditions}: Especially for IoT devices, the surrounding physical conditions can have a significant impact on operations. Parameters like temperature, humidity, air quality, and light conditions can influence device performance and decision-making. 
        \item\textbf{Location Specifics}: Beyond just knowing 'Where am I?', understanding the specifics of that location, such as whether it is indoors or outdoors, urban, or rural, can provide valuable context for system operations. This knowledge can help tailor system behavior to local conditions, enhancing adaptability and efficiency.
        \end{itemize}
        \end{itemize}
    
    The context, as defined across these dimensions, provides a holistic framework to inform design decisions, guides system behavior, and ensures that the Cognitive Internet is responsive, adaptive, and intelligent in its operations.
    \subsection{Dismantling Application Silos for Contextual Software Defined Systems}

    One of the most notable characteristics of Cognitive Internet is its capacity to dismantle application silos and usher in dynamic and interconnected communication among devices. Let us illustrate this with a practical example: consider two smart home devices—a smart thermostat and a smart door lock. In the earlier era of IoT, these devices operated within isolated silos, each performing their predefined function independently. They could eventually be synchronized by a cloud-based orchestration, but this necessitates the devices to send data to the cloud.
    
However, in the Cognitive Internet era, a profound shift occurs. These devices actively discover, communicate, and collaborate directly with each other based on context. It is important to note that these devices are not considered or seen as isolated devices attached to a network as external entities, but they inherently belong to the same context (be it a subnetwork, locale, or account), a fact known by every device. As a result, each device can judiciously decide to relay information to other devices within the same context. This explicit way of operation avoids the unwanted side effects of publishing events blindly and developing applications for a fixed pre-defined operation. Instead, systems expand and shrink as per based the context at the time therefore they are Contextual Software Defined Systems vs. fixed applications. For instance, when the smart door lock is engaged, signaling the homeowner's departure, it seamlessly and consciously relays this information to the smart thermostat. In response, the thermostat intelligently adjusts the home's temperature settings to conserve energy while ensuring optimal comfort.

This level of direct cooperative decision-making and communication is a hallmark of the Cognitive Internet. It is where devices move beyond their individual roles, working in concert to make context-aware, autonomous decisions.

By gaining a deeper understanding of the definition, characteristics, and integral components of the Cognitive Internet, businesses can effectively navigate the challenges and harness the vast opportunities that this digital revolution presents. In the subsequent section, we will delve further into these challenges while highlighting the imperative need for a novel infrastructure that can support and sustain this remarkable transformation.
\subsection{Understanding Knowledge-as-a-Service (KaaS) }
In the Cognitive Internet paradigm, the interplay between devices, microservices, and context is of paramount importance. Each microservice, running on distinct devices, is not merely an executor of pre-defined tasks. Rather, they function as knowledge-gatherers, extracting data from their environment and converting it into actionable knowledge.

\textbf{Knowledge Extraction and Conversion}: At the heart of this ecosystem lies the process of transforming raw data into knowledge. Devices in the Cognitive Internet are embedded with microservices that continuously observe, learn, and interpret data from their surroundings. This data, when processed and analyzed, becomes knowledge, a more refined and valuable asset. Knowledge, in this sense, goes beyond mere facts—it represents an understanding, a discernment that can be acted upon.

\textbf{Knowledge Exchange and Financial Transactions}: KaaS facilitates the exchange of this knowledge among services. Just as Software-as-a-Service revolutionized the distribution of software, KaaS is paving the way for a seamless transfer of knowledge. Services can request, share, or sell knowledge to other services within the Cognitive Internet framework. This exchange is not only restricted to knowledge; financial transactions can also be integrated, providing an incentive for services to generate and share valuable insights. For instance, a service that has perfected energy conservation in smart homes can share this knowledge with others for a fee.

\textbf{Ranking and Feedback Mechanisms}: The efficacy of these knowledge transactions is upheld by a robust feedback and ranking system. Services can rate each other's performance based on the outcomes of the shared knowledge. This continuous feedback loop ensures that only the most accurate and valuable knowledge is circulated. Over time, services that consistently provide high-quality knowledge can be recognized and prioritized, while those with less accurate insights might be flagged for improvement.

\textbf{Incorporating Context}: Context plays an indispensable role in KaaS within the Cognitive Internet. By understanding the context in which they operate, services can fine-tune the knowledge they produce and consume. For instance, knowledge relevant to a home setting might differ from that of an industrial environment. Context-aware microservices can discern these nuances and adapt accordingly. Context also plays a significant role in focusing the limited device resources, especially on edge, towards knowledge that might be required for the situation only.

In summary, KaaS in the Cognitive Internet ecosystem represents a leap from static, siloed operations to a dynamic, collaborative, and context-aware environment. This model allows services to extract and share knowledge, while also introducing mechanisms for ranking and scoring services based on measured outcomes. The integration of financial transactions further monetizes the exchange of valuable insights, fostering a culture of shared learning and mutual growth. With both financial incentives and performance-based rankings, services are encouraged to continually elevate their efficacy and collaboration, ensuring a thriving and evolving ecosystem.

\section{TRANSFORMING INTELLIGENT SERVICE ECOSYSTEMS}
In the age of the Cognitive Internet, a profound transformation is sweeping through the realm of digital services. We are witnessing a departure from isolated applications and conventional Software as a Service (SaaS) models toward a dynamic ecosystem of intelligent services that possess adaptability, responsiveness, and deep interconnectivity.
\subsection{Beyond Isolated Apps and Traditional SaaS Models}
Traditionally, applications and services functioned in isolation, and even SaaS models, while fostering connectivity, often retained a siloed approach. Data and insights remained confined within the boundaries of individual services, limiting the potential for genuinely dynamic, context-aware operations.
However, the Cognitive Internet era demands a radical departure from these static and isolated models. The sheer volume, speed, and diversity of data generated in this era necessitate services that not only process information swiftly but also share their insights and collaborate seamlessly to drive informed and coordinated actions.
\subsection{Cognitive Services at the Heart of any Intelligent Service Ecosystem}
At the core of this Intelligent Service Ecosystem lies the concept of cognitive services. These services represent a new generation of digital offerings that transcend the mere execution of predefined tasks. They are distinguished by their intelligence and adaptability, harnessing AI algorithms to learn from data, refine their operations, and autonomously make decisions.

Within the Intelligent Service Ecosystem, cognitive services function as the building blocks. They can operate independently, each offering a unique array of capabilities. Yet, they are intrinsically designed to collaborate and complement each other, resulting in an agile, flexible, and dynamic ecosystem. Here, services can seamlessly combine to tackle a diverse range of tasks.

\subsection{Collective Knowledge Sharing and Adaptation}
The unique aspect of the Intelligent Service Ecosystem lies in how these services discover and adapt to tasks. It transcends the boundaries of individual services. Each service learns from its experiences and continually enhances its performance. However, the true innovation emerges from their collective learning.

When a service discovers an innovative approach to accomplishing a task, it shares this newfound knowledge across the ecosystem. This collaborative learning gives rise to an ever-evolving ecosystem, where enhancements in one service catalyze advancements in others.

Another form of adaptation is seen in the communication dynamics between services operating within the same context. A deep understanding of their operating context enables a service to establish dialogues with other services, agreeing on semantic protocols for efficient collaboration. Historically, common ontologies and intricate semantic annotations of data have facilitated this. In the Cognitive Internet realm, this goal is further refined, allowing services to directly communicate and converge on shared semantics. This mutual understanding and agreement are fostered by their shared contextual backdrop, making the inter-service communication more coherent and effective.

Consequently, the Intelligent Service Ecosystem represents an intricately connected network of learning and evolving cognitive services. They operate in harmony, collectively harnessing their intelligence to adapt to tasks more efficiently. This marks a significant departure from the isolated operations of the past, as it leverages the collective knowledge and wisdom of diverse services to optimize outcomes in the Cognitive Internet era. In this paradigm, services form and act based on collective knowledge rather than isolated and siloed interfaces, all while retaining their autonomy.
\section{HYBRID EDGE CLOUD: AN INDISPENSABLE ENABLER FOR COGNITIVE INTERNET}
In the Cognitive Internet era, end devices are not just data collectors but autonomous cognitive agents that operate akin to human intelligence. As shown in Figure 4, HEC is a fusion of edge and cloud computing, and the driving force behind this transformation, empowering a distributed AI infrastructure supporting the Internet of Systems. It enables dynamic, real-time, and intelligent services emblematic of Cognitive Internet.
\begin{figure}[ht]
    \centering
    \includegraphics[width=.6\linewidth]{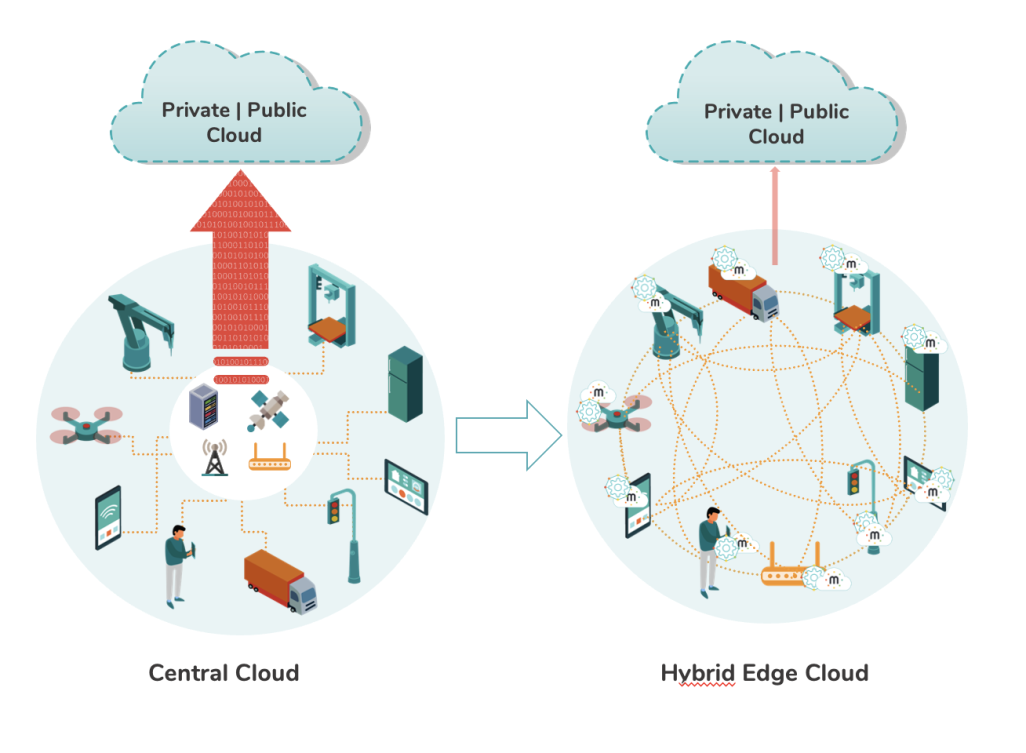}
    \caption{High level comparison of central cloud and hybrid edge cloud.}
\end{figure}
\subsection{Rethinking Edge Computing: Beyond the Network Edge}
Traditionally, "edge" denoted the network edge, where data was collected and transmitted. However, in the Cognitive Internet era, "edge" encompasses end devices—smartphones, machines, robots, and more—that possess autonomous decision-making capabilities. They perceive, learn, decide, and act upon information, becoming autonomous and collaborative cognitive agents.
\subsection{Hybrid Edge Cloud: Balancing Autonomy and Global Learning}
HEC bridges the edge and cloud computing realms, offering the best of both worlds. At the edge, data processing occurs close to the source using microservices that can adapt to the needs of the solution, reducing latency, conserving bandwidth, and enabling real-time decision-making. This proximity to the source is essential for applications demanding rapid responses, such as autonomous operations, industrial automation, and augmented reality.

Simultaneously, HEC maintains connections to central cloud resources. This enables global learning, coordination, and long-term insights. The distributed AI infrastructure seamlessly integrates with the global cloud while ensuring local autonomy for end devices.
\subsection{Benefits of the Hybrid Edge Cloud}
As we navigate into the era of the Cognitive Internet—a paradigm that envisions an intelligent, adaptive, and highly interconnected digital ecosystem—Hybrid Edge Cloud (HEC) emerges as a foundational infrastructure, powering intelligence and bestowing autonomy to end devices. The benefits HEC offers are not only transformative for current technologies but are also instrumental in shaping the trajectory of Cognitive Internet.

HEC introduces profound benefits in business operations, aligning them with the dynamic nature of Cognitive Internet and its emerging business models. From a technical, architectural, and operational standpoint, HEC ensures financial efficiencies, addressing the rapid and evolving demands of this new era. Furthermore, in the software development sphere, HEC refines and streamlines processes. It champions an abstraction layer that not only accelerates time-to-market but also promotes a shift from standalone applications to agile, reusable services, aligned with the seamless adaptability required for Cognitive Internet.

From a technology vantage point, the adoption of HEC yields numerous transformative advantages:
\begin{itemize}
    \item \textbf{Scalability}: HEC ensures that the cloud infrastructure grows alongside the expanding network of intelligent devices. Unlike a centralized cost-based approach treating devices as clients, HEC allows cloud resources to expand as more end devices with computing resources are added.
    \item \textbf{Reduced Latency}: Real-time decision-making, crucial for applications like autonomous vehicles and robotics, hinges on low latency. By processing data at the source, latency is minimized, enhancing the performance of time-critical systems.
    \item \textbf{Bandwidth Efficiency}: The Cognitive Internet generates vast volumes of data. Transmitting all this data to central cloud systems or intermediary gateways is neither efficient nor sustainable. HEC optimizes bandwidth by processing data locally, sending only essential insights and updates to the central cloud.
    \item \textbf{Cost Efficiency}: HEC achieves cost efficiency by optimally offloading workflows to end devices, significantly reducing the burden on centralized cloud infrastructure and, in turn, lowering cloud hosting costs for businesses.
    \item \textbf{Enhanced Privacy}: With HEC, services can process data locally and decide on where to store knowledge and who to share it with, based on the privacy policy of each transaction. This reduces the risk of data breaches and ensuring compliance with stringent privacy regulations.
    \item \textbf{Enhanced Security}: By deploying an API gateway at the edge along with a zero-trust security approach, HEC enables safeguarding against unauthorized intrusions, maintaining data integrity throughout its lifecycle.
    \item \textbf{Local Decision-Making Autonomy}: This is a fundamental advantage of HEC. In situations where real-time decision-making is critical, HEC allows end devices to autonomously make immediate decisions. This independence from centralized cloud resources ensures that real-time decisions can occur at the edge, while also maintaining the option to leverage cloud orchestration for further refinement based on edge context. This balance between local and global decision-making introduces the concept of an "eventually optimal system."
    \item \textbf{nergy Efficiency}E: In today's sustainability-focused landscape, managing energy consumption is of paramount importance. HEC contributes to energy efficiency by processing data on-site, which curtails the need for extensive energy waste during data transmission, lessens the demand on energy for network and routing operations to the central cloud, and reduces the overall energy usage of cloud processing
    \item \textbf{Zero Configuration}: HEC stands out for its adaptability and intuitive operation, significantly reducing the need for system and device pre-configuration. This streamlined approach ensures swift deployment. In contrast to traditional cloud orchestration, which relies on predefined event-based workflows and resource allocation, HEC adopts a choreography model at the edge. This approach enables dynamic coordination and communication among nodes, leveraging contextual information and situational awareness. By processing data at the edge, HEC promotes a more flexible and responsive network topology. Nodes possess the intelligence to make real-time decisions, fostering an agile environment where configurations adapt to changing conditions without the constraints of cloud-based orchestration.
    \item \textbf{Interoperability}: Within the diverse landscape of the Cognitive Internet, HEC facilitates seamless microservice-level interactions among various devices, operating systems, platforms, services, and networks, promoting an integrated environment. This interoperability is achieved through well-defined APIs for each HEC service, outlining their offered resources and functions. Additionally, HEC employs service discovery mechanisms, enhancing the visibility and dynamic engagement of these services. This dual strategy enables efficient communication between diverse services and allows them to adapt to each other's capabilities, resulting in a cohesive and collaborative network of resources.
    \item \textbf{Contextual Awareness}: HEC empowers services to sense, comprehend, and respond to their immediate environment, enabling personalized and pertinent actions. It exposes information such as ownership, capabilities, the serving application or service, the device’s role, location, and its relationships with other entities, among other contextual factors.
    \item \textbf{Device Autonomy and Collaboration}: HEC grants end devices the autonomy for on-the-spot decisions, and yet the opportunity to collaborate dynamically with other autonomous entities and based on context that reduces dependency on distant cloud resources.
    \item \textbf{Device Adaptability}: HEC ensures that system on end devices can swiftly adapt to changing conditions using a microservice-based approach where device capabilities can be updated dynamically based on real-time context.
    \item  \textbf{Agile Transformation of Applications}: HEC can evolve traditionally siloed applications into modular, task-specific services. These services, instead of operating in isolation or being perpetually dependent on central cloud, can interact directly with one another, allowing for more versatile and adaptive solutions in near applications that require real-time and near real-time applications. By minimizing dependency on the central cloud, HEC paves the way for services to collaboratively execute complex tasks, swiftly respond to ever-changing demands, and ensure a consistently high-quality user experience, regardless of the central cloud accessibility.
    \item \textbf{KaaS (Knowledge as a Service) Enablement}: By bridging the edge with the cloud, HEC acts as the backbone for the KaaS model. This facilitates a dynamic exchange of knowledge between services, both amongst edge nodes and with the central cloud, benefiting businesses substantially.
    \item \textbf{Data Brokerage} : Using HEC, individual devices or expansive systems can treat their data as a tangible asset and expose the asset to others to monetize or trade.
    \end{itemize}
    
To streamline the development and deployment of applications for the Cognitive Internet, HEC offers the following key benefits:
\begin{itemize}
    \item \textbf{Device, OS, and Network Abstraction}: HEC provides abstraction layers for device hardware, operating systems, and network compatibility, creating a standardized environment where services can be developed once and seamlessly deployed across diverse settings without the need for extensive modifications.
    \item \textbf{Accelerated Software Delivery}: Through this abstraction, HEC reduces software fragmentation and promotes a consistent software development approach, simplifying deployment methods and resulting in faster software delivery and iteration cycles.
    \item \textbf{Efficient Software Asset Management}: HEC encourages the creation of solutions composed of a combination of services, leading to fewer digital assets to manage and maintain. By reducing overhead and allowing teams to focus on optimizing a smaller pool of assets, HEC ensures the quality and efficiency of each service
\end{itemize} 

The combined impact of these advantages is twofold. Firstly, businesses experience a tangible reduction in development costs. Secondly, there is a paradigm shift in how services are conceived and utilized. Rather than continually creating applications for various tasks and domains, the focus shifts to the unique value each service offers. This value-centric approach optimizes resource utilization and enhances the overall effectiveness of the digital ecosystem.
\section{CASE STUDIES}

As we delve into these real-world scenarios, it becomes evident how devices not only become intelligent, capable of hosting microservices and APIs, but also how they cultivate a self-awareness, understanding the context in which they operate. As opposed to for example, traditional models, particularly in IoT, primarily emphasized devices reporting to a global authority. However, the advent of the Cognitive Internet era amplifies this foundation by integrating intelligence and context-awareness alongside this global connectivity. These case studies underscore the importance of these three properties, illustrating a shift in the technological paradigm.

\subsection{Software-Defined Vehicles: Navigating the Cognitive Internet with Precision}
The automotive industry is rapidly evolving, with the advent of software defined, driver-assisted  and autonomous vehicles at the forefront of this transformation. These vehicles are equipped with an array of sensors, cameras, Lidar, and advanced AI systems, all generating massive amounts of real-time data and are also enhanced with ad hoc devices like the passenger's phones, the traffic light of a smart city or the different devices or the smart house.

In this context, as illustrated in Figure 5, Cognitive internet using HEC as a game-changer. These modern vehicles require split-second decision-making for navigation, obstacle avoidance, and ensuring passenger safety. By processing critical data locally, these vehicles can make immediate decisions, enhancing safety and efficiency. Autonomous vehicles are becoming part of a dynamic system that expand and shrink dynamically based on the context in which they reside.

Simultaneously, HEC connects these vehicles to a global network. Data from millions of software defined vehicles can be aggregated, analyzed, and employed to train AI models for continuous improvement. This balance between local decision-making and global learning epitomizes the power of the Cognitive Internet.

\begin{figure}[ht]
    \centering
    \includegraphics[width=0.6\linewidth]{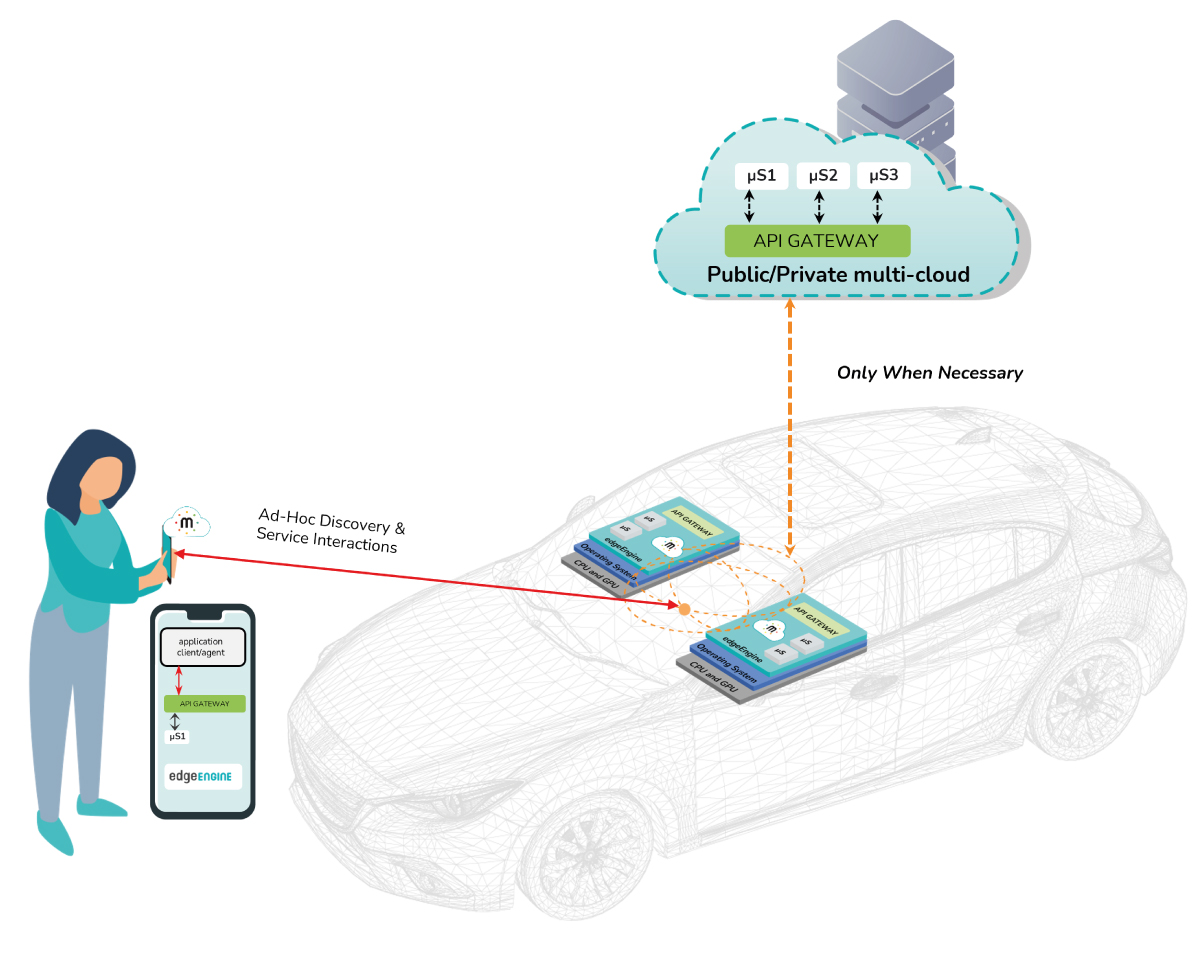}
    \caption{Software Defined Vehicles with HEC.}
\end{figure}

\subsection{Industry 5.0: Collaborative Manufacturing with Cognitive Systems}
In the realm of Industry 5.0, manufacturing processes are undergoing a radical transformation. Cognitive systems powered by the Cognitive Internet are driving collaborative manufacturing to new heights of efficiency and adaptability.
Factory floors are now populated with smart machines, robots, and sensors that operate as autonomous cognitive agents. These agents not only carry out manufacturing tasks but also make real-time decisions based on their local insights. For example, if a robotic arm detects a defect in a product, it can immediately adjust its actions to rectify the issue, ensuring high-quality production.

Furthermore, these cognitive agents collaborate with one another, forming an interconnected service mesh of intelligence. Machines share knowledge, optimize production schedules, and adapt to changing requirements collaboratively. This collaborative manufacturing approach reduces downtime, improves product quality, and enhances overall efficiency.

As shown in Figure 6, HEC plays a pivotal role in this Industry 5.0 transformation. It enables local decision-making autonomy for machines and robots while facilitating global learning and coordination. Data generated by machines across multiple manufacturing sites can be analyzed centrally to identify trends, optimize processes, and ensure consistent product quality.

In essence, Industry 5.0 leverages the Cognitive Internet to create intelligent, interconnected manufacturing ecosystems that prioritize efficiency, adaptability, and collaboration among machines and systems. It exemplifies how the Cognitive Internet, coupled with HEC is reshaping industries for a more intelligent and responsive future.
\begin{figure}[ht]
    \centering
    \includegraphics[width=0.6\linewidth]{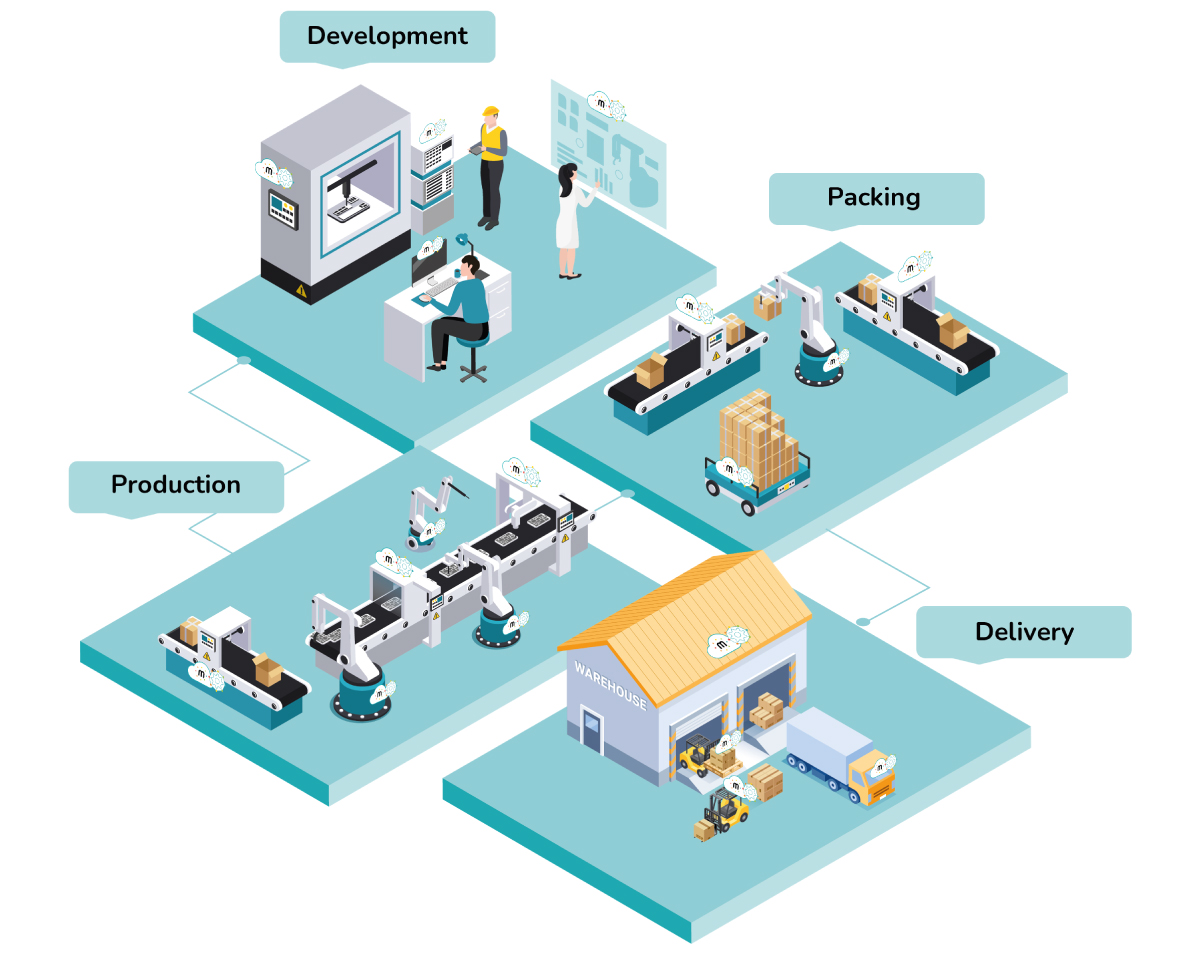}
    \caption{Industry 5.0 with HEC.}
\end{figure}
\subsection{Healthcare: Revolutionizing Patient Care with Intelligent Devices}
The healthcare sector is embracing the Cognitive Internet to revolutionize patient care. Smart medical devices, wearables, and monitoring equipment now possess the capability to interpret patient data, identify anomalies, and trigger alerts in real time.

Consider a patient's wearable device that continuously monitors vital signs as illustrated in Figure 7. If a critical deviation is detected, the device can autonomously trigger an alert, potentially saving lives and this in the context in which it resides. For example, the alert can therefore be sent to persons passing by with a cognitive fitness watch to draw attention to a medical condition that requires immediate attention while preserving their privacy. Simultaneously, this data can be shared with healthcare providers ensuring immediate medical intervention.

Moreover, HEC supports local storage and processing of sensitive patient data, mitigating privacy concerns. This intersection of real-time decision-making, data privacy, and global connectivity illustrates the profound impact of the Cognitive Internet on healthcare.
\begin{figure}[ht]
    \centering
    \includegraphics[width=0.6\linewidth]{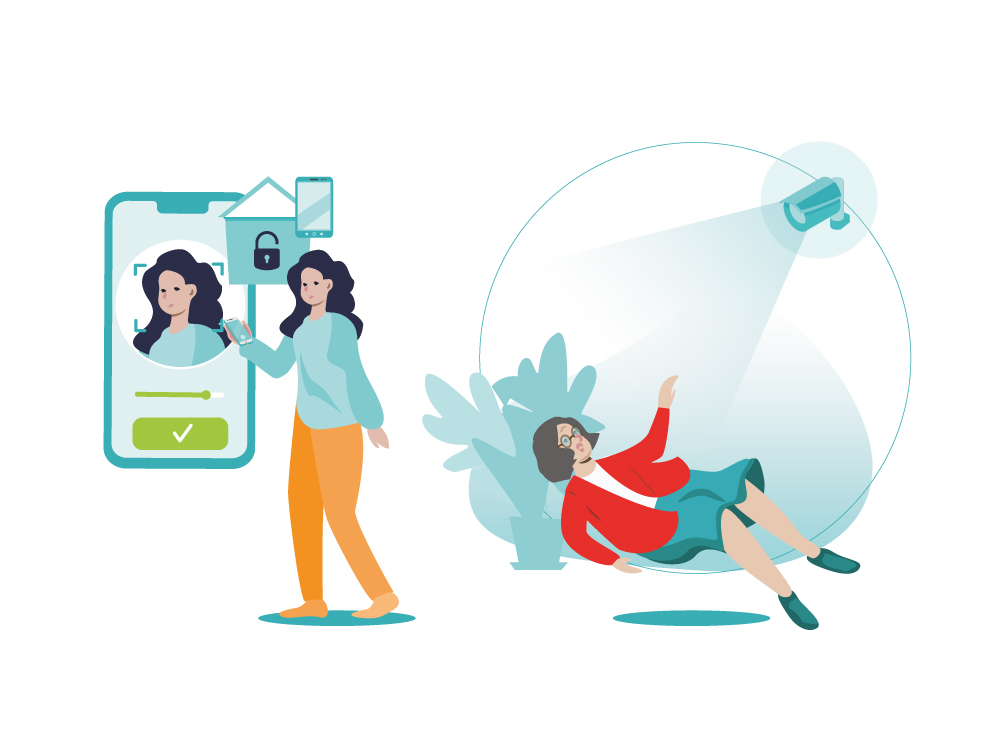}
    \caption{Digital wellness and healthcare with HEC.}
\end{figure}
\section{FUTURE PERSPECTIVES AND SOLUTIONS }
As the Cognitive Internet matures and expands its reach, we stand on the precipice of a transformative shift in how we perceive and interact with digital systems. The future no longer holds a multitude of independent systems interacting in often cumbersome and inefficient ways. Instead, we envision a unified, context-driven system, a kind of digital organism, which dynamically adjusts its scale and function according to the specific needs and situations it encounters.

The profound implication of this perspective is that the Cognitive Internet, rather than merely being a tool for interconnecting devices and processes, becomes a seamless fabric that understands and reacts to its environment. Think of it less as a network of individual components and more as a fluid, adaptable entity that embodies the concept of context at its very core.

For industries and individuals alike, this offers unparalleled advantages:
\begin{itemize}
    \item \textbf{Responsiveness}: The system can instantly resize, redirect, and repurpose resources based on real-time demands.
    \item \textbf{Efficiency}: A singular context-aware system eliminates redundancies, streamlining operations and reducing waste.
    \item \textbf{Intelligence}: By operating as a cohesive unit, the system can harness collective insights, making smarter decisions faster than a disjointed network of entities.
    \item \textbf{Scalability} : As needs grow or diminish, the system can effortlessly expand or contract, ensuring optimal performance without unnecessary overheads.
\end{itemize}
In this new era, our approach to design, integration, and optimization will fundamentally change. We won't be piecing together separate systems; we'll be shaping a singular, malleable entity that's tuned to the nuances of its context. It's an exciting horizon, and as we delve deeper into this section, we'll explore the myriad of possibilities and solutions that this revolutionary perspective brings to the fore.

\subsection{Smart Cities}
In smart city initiatives, the Cognitive Internet can enhance urban planning, traffic management, energy efficiency, and public safety. Autonomous systems, powered by HEC, can optimize traffic flow, monitor environmental conditions, and respond to emergencies in real time.
\subsection{Agriculture}
Precision agriculture benefits from the Cognitive Internet by optimizing crop management, resource allocation, and environmental monitoring. Smart sensors, drones, and autonomous farming equipment leverage local intelligence while benefiting from global insights to enhance crop yield and sustainability.
\subsection{Industrial Automation}
Manufacturing and industrial processes increasingly rely on intelligent automation. Robots, machines, and sensors can collaboratively optimize production, predict maintenance needs, and minimize downtime through real-time decisions and global learning.
\subsection{Healthcare}
In telemedicine and remote patient monitoring, the Cognitive Internet ensures timely interventions and data privacy. Wearable devices and medical equipment can autonomously detect health issues, share data with healthcare providers, and enable remote consultations.
\subsubsection{Environmental Monitoring}
Real-time data collection and analysis support environmental conservation efforts. Drones, sensors, and autonomous monitoring stations equipped with AI can identify pollution sources, monitor wildlife habitats, and respond to ecological emergencies.
\section{CONCLUSION}
The era of the Cognitive Internet is upon us, heralding a transformative shift in how we perceive, interact with, and harness intelligent systems. Cognitive Internet inherently infuses intelligence both horizontally and vertically throughout the entire network allowing to fuse organically and fully the realm of cognitive IoT and human intelligence. This inherent intelligence empowers cognitive choreography, facilitating interactions between connected devices, services, entities, and individuals across diverse domains and industries. Importantly, it upholds the autonomy of decision-making and accommodates heterogeneous identities.

In this new digital epoch, devices cease to be mere conduits of data; they become self-aware cognitive agents capable of autonomous decision-making and dynamic collaboration. The Intelligent Service Ecosystem thrives on collective knowledge sharing and adaptation, fostering a network of cognitive services that continually refine their operations based on real-world experiences. This collaborative paradigm replaces isolated application silos with a dynamic, interconnected digital landscape.

Central to this transformation is HEC a platform that empowers a distributed AI infrastructure that balances local autonomy with global learning. It delivers scalability, reduced latency, bandwidth efficiency, enhanced privacy, and energy efficiency, ensuring that the Cognitive Internet operates seamlessly and efficiently.

As the Cognitive Internet continues to evolve, its applications span diverse domains, from smart cities and agriculture to industrial automation and healthcare. These applications exemplify the real-world impact of the Cognitive Internet, where autonomous devices and systems make decisions, learn from experience, and drive an era of unprecedented connectivity and intelligence.

In embracing the Cognitive Internet, enterprises must recognize the imperative of deploying a modern distributed cloud and AI infrastructure that empowers devices and systems to thrive in this new digital era. The path forward demands a deep understanding of the Cognitive Internet's characteristics, challenges, and potential, as well as a commitment to fostering the sustainable, intelligent, and interconnected world that it promises to create.

\bibliographystyle{plain}
\bibliography{bibfile}


\end{document}